\documentclass[a4paper,11pt]{article}
\usepackage{pos}

\usepackage{siunitx}
\usepackage{graphicx}

\graphicspath{{figures/}{../ichep_ship_talk_matei/graphics/}}

\title{The detector system of the SHiP/NA67 experiment at CERN}

\author*[a]{Matei Climescu}
\onbehalf{on behalf of the SHiP Collaboration}

\affiliation[a]{Department of Physics and Astronomy, Ghent University,\\
  Proeftuinstraat 86, B-9000 Ghent, Belgium}

\emailAdd{matclim@cern.ch}

\abstract{%
The SHiP/NA67 experiment aims to search for feebly interacting GeV-scale new particles and to
perform all-flavour neutrino-physics measurements at the HI-ECN3 beam facility at the CERN SPS.
The collaboration is currently optimising the experiment's initial configuration for the
commissioning and first physics runs of 2032--2033. The detector subsystems comprise a few
large-area instruments: for new-particle searches, highly sophisticated veto detectors, high
timing-resolution detectors, a lightweight straw tracker system, and high-spatial-precision
calorimeters; and, for neutrino reconstruction, small transverse-size, high-granularity
calorimeters.
}

\FullConference{43rd International Conference on High Energy Physics (ICHEP 2026)\\
30 July  to 5 August , 2026\\
Natal, Brazil\\}

\begin{document}
\maketitle

\section{Introduction}

New physics is required to explain anomalies in our understanding of the universe and has thus far escaped
detection for one of two reasons. It may be \emph{too heavy} to have been produced at existing
machines, which is the case addressed by the energy frontier~\cite{briefingbook}. Alternatively
it may be \emph{too weakly coupled}: light states whose interactions with the SM are far feebler
than those of neutrinos would have evaded every search performed so far not because of their
mass but because of their production rate. Such feebly interacting particles (FIPs) arise
naturally, for instance in hidden-sector constructions, where communication with the SM proceeds
through \emph{portals}: vector, scalar, neutrino or axion-like particles~\cite{physicscase}.

This second range of possibilities constitutes the intensity frontier: the suppression in
coupling must be compensated by a very large number of produced particles. This is the direction
taken by SHiP/NA67~\cite{proposal,ship2025}, located at the North Area's Beam Dump Facility (BDF)
approved by CERN in March 2024. The experiment, being designed for the breaching of the intensity
frontier, makes use of an innovative detector system. Each subsystem described below exists
because of specific requirements, which are typically coupled.

\section{Experimental concept and detector requirements}
\label{sec:concept}

A FIP coupled to the SM inherits a decay width proportional to the square of its small coupling,
which for the masses and couplings of interest places its laboratory decay length in the range
$c\tau \sim \mathcal{O}(\SI{e-3}{}) - \mathcal{O}(\SI{e3}{})\,\si{\meter}$. The experiment that
follows must combine a high-intensity beam, a long decay volume, precision scattering and decay
detectors, and sophisticated background suppression. SHiP will receive
$\mathcal{O}(\SI{4e19}{})$ protons on target (PoT) per year at \SI{400}{\giga\electronvolt}/$c$,
accumulating \SI{6e20}{} PoT over the programme; this yields $\mathcal{O}(10^{18})$ charmed and
$\mathcal{O}(10^{16})$ beauty hadrons, the decay products of which are then given \SI{50}{\meter} of baseline in which
to decay back into visible SM particles.

Neutrino and muon deep-inelastic scattering, as well as muon
combinatorial are the main backgrounds. They are brought down using the SHiP infrastructure and detector
system, with the combination reducing them far below one event in fifteen years.

\begin{figure}[htbp]
  \centering
  \includegraphics[width=0.88\textwidth,height=6cm,keepaspectratio]{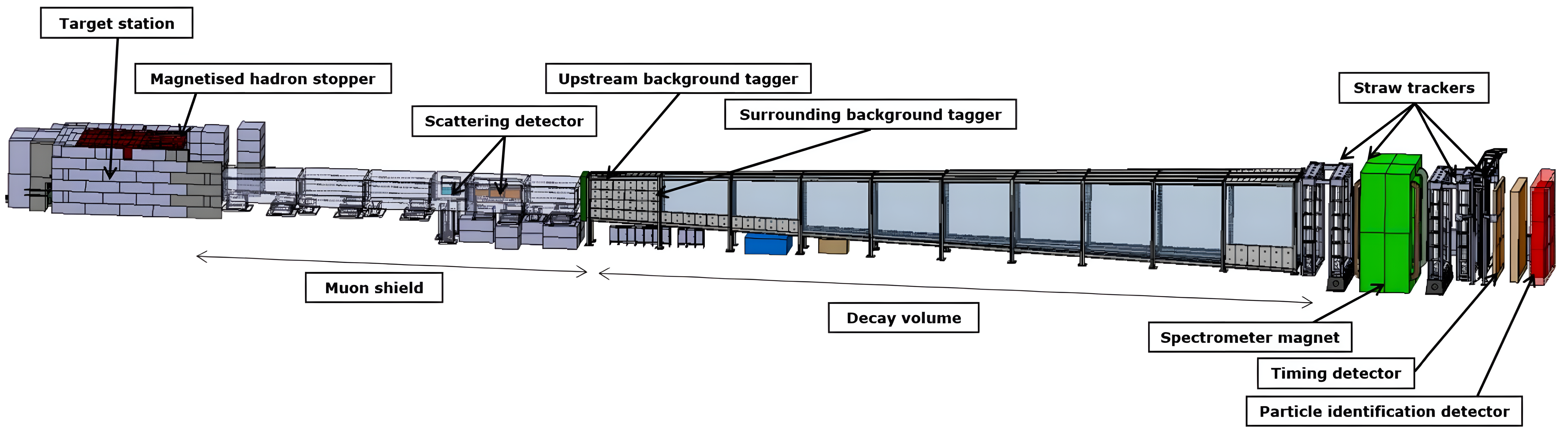}
  \caption{Layout of the BDF/SHiP experiment in the ECN3 hall. The proton beam enters from the
  left onto the tungsten target; the active muon shield deflects the residual muon flux while
  housing a scattering detector; the background taggers surround the helium-filled decay volume,
  upstream of the spectrometer, timing detector and calorimeter system.}
  \label{fig:layout}
\end{figure}

A further requirement is that SHiP's data acquisition is fully triggerless. An SPS spill at the
BDF will last \SI{1}{\second}, and the experiment records the entire duration; there is as such
no notion of an ``event'' at readout level, only of time windows reconstructed offline. Every
subsystem is thus free-running and self-timestamping. They are shown together with the facility in
Fig.~\ref{fig:layout}.

\section{Upstream systems}
\label{sec:upstream}

\subsection{Target and target complex}

The target is built from pure tungsten, chosen to maximise the heavy-flavour production cross
section per unit length and to minimise induced muon and neutrino background. It receives
\SI{4e13}{} protons per spill, which imposes active cooling; a slotted geometry through which coolant is circulated between the tungsten blocks is thus adopted. High-pressure gaseous Helium cooling
is the current baseline and has been demonstrated experimentally. A magnetised hadron
stopper is located at the rear of the target complex.

Several target prototypes were built and tested during 2025 addressing the cooling strategy, the mechanical integrity of the target and its beam-monitoring
instrumentation under repeated thermal shock with dedicated stress tests performed. The radioactive waste management strategy for the irradiated target has a direct impact on the permissible target length. Following this campaign
the target design is now essentially frozen, with the Technical Design Report (TDR) for the facility being due by the end of 2026.

\subsection{Active muon shield}

SHiP, in order to filter the intense muon flux emerging from the target complex, employs an \emph{active} muon shield: a chain of magnets
which sweep muons out of the acceptance before they reach the decay volume. The design adopted is
the custom-developed \textbf{Tuned Return Yoke} (TRY) together with a downstream reversed polarity
concept~\cite{mushield}, which reduces the flux from $\mathcal{O}(10^{11})$ to
$\mathcal{O}(10^{5})$ per spill. High-momentum muons are deflected in the upstream section, where the available field
integral is largest relative to their rigidity. Low-momentum muons are absorbed in the yoke
material. Intermediate-momentum muons are mitigated by an optimised dilution of the field.
Residual trajectories are finally removed by a downstream section of reversed polarity. 
A modular muon shield has been chosen, that allows inserting detectors in the return yoke and, if required, a reconfiguration of the return field distribution.

\section{The Scattering Detector}
\label{sec:sd}

The all-flavour neutrino flux produced in the dump cannot be swept away. Rather than treat it
purely as a background, SHiP will study it: the Scattering Detector (SD) is placed inside the
muon shield, enabling the magnetic field to be utilised for reconstruction of final-state momenta.

\begin{figure}[htbp]
  \centering
  \includegraphics[width=0.88\textwidth,height=6cm,keepaspectratio]{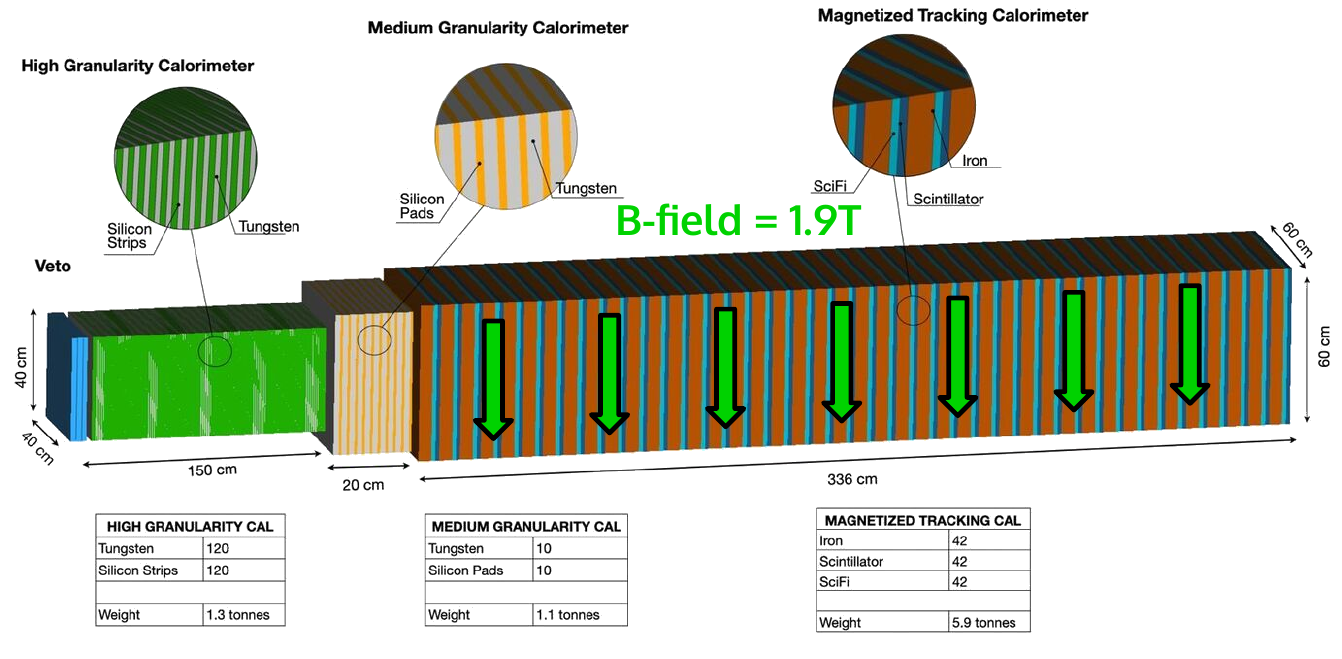}
  \caption{Layout of the SHiP Scattering Detector.}
  \label{fig:sd}
\end{figure}

The scattering detector in particular proposes to bring forward the era of $\nu_\tau$ phenomenology through its expected $\mathcal{O}(10^4)$ interactions against the current  $\mathcal{O}(10^1)$ candidates today with even larger samples of the
other flavours. In addition, light dark matter scattering off atomic electrons can be accessed
within the SD. The topological reconstruction of these signatures and the smallness of the cross
sections drive the design of the detector, yielding a long, multi-section and highly granular
detector operating in tandem to allow optimal reconstruction efficiency. The detector is thus
composed of a plastic scintillator-based veto detector followed by a high-granularity calorimeter
utilising silicon strips recycled from the CMS Tracker Outer Barrel, silicon pixels and tungsten
as a passive material. It precedes a medium-granularity calorimeter composed of silicon pads and
tungsten, which itself is followed by a magnetised tracking calorimeter composed of scintillating
fibres affording it tracking capabilities, plastic scintillator tiles granting it energy
resolution and iron as a passive material. The detector can be visualised in Fig.~\ref{fig:sd}.

\section{Achieving a background-free search}
\label{sec:taggers}

\subsection{Decay vessel}

The decay volume is the \SI{50}{\meter} of helium-filled space in which a FIP may decay, designed
to minimise $\nu$ and $\mu$ inelastic scattering. The helium will be contained in a single
$<\SI{0.3}{\milli\meter}$ thick polymer balloon and utilise a helium purification and
recirculation system, whose cost will be amortised after a single year of operation.

\subsection{Upstream Background Tagger}

The Upstream Background Tagger (UBT) is a window at the entrance of the decay volume that tags
muons entering the fiducial volume, so that any downstream vertex coincident with an incoming
track can be vetoed. It must cope with a highly non-uniform rate across the plane, with greater
muon fluxes on the sides as a result of the muon shield's sweeping. The adopted solution combines
two technologies according to local rate: straw tubes in the low-rate regions, where their
spatial resolution is exploited, and scintillator tiles of different sizes across the full plane.
Optimisation of the system is ongoing, with the limiting factor in the hot regions being the
readout electronics. The FastIC+ ASIC~\cite{fastic} is the current baseline and its rate
capability sets the achievable segmentation.

\subsection{Surrounding Background Tagger}

The Surrounding Background Tagger (SBT) encloses the decay volume along its full length. It tags
charged particles entering from outside and, crucially, neutrino and muon interactions occurring
in the helium as well as the vessel wall itself --- the irreducible component that cannot be
removed upstream.

The SBT comprises 780 compartments filled with a total of $\sim\SI{145000}{\liter}$ of linear
alkylbenzene-based liquid scintillator. Each cell is read out by wavelength-shifting optical
modules (WOMs) coupled to SiPMs, with 40 SiPMs per module. The WOM approach is well matched to
the geometry: it collects light over a large area and concentrates it onto a small photosensor
area, keeping the channel count and the dark-count rate tractable for a detector of this volume.
The design achieves better than \SI{99}{\percent} tagging efficiency with a timing resolution of
$\sim\SI{1}{\nano\second}$.

An extensive test-beam campaign is in progress to consolidate the design. It addresses the
performance of the FastIC+ readout chain; edge effects at the cell walls, which set the efficiency
in the least favourable part of each cell and therefore of the system as a whole; the achievable
position and timing resolution; the operation of liquid-scintillator cells at this scale; and
their long-term stability over a fifteen-year experiment. Taken together with the UBT and with
simple kinematic and impact-parameter selections, these systems reduce the dominant backgrounds
to a negligible level while preserving signal efficiency.

\section{The Hidden Sector Decay Spectrometer}
\label{sec:hsds}

\subsection{Straw spectrometer and magnet}

The spectrometer consists of four straw tracker stations coupled to an energy-efficient
superferric dipole magnet built with an MgB$_2$ superconducting coil, providing a nominal on-axis
field of \SI{0.15}{\tesla} and a bending power of \SIrange{0.6}{0.8}{\tesla\meter}. The sensitive
area is $4\times\SI{6}{\meter\squared}$ and the rate is $<\mathcal{O}(\SI{10}{\kilo\hertz})$. Each station uses straws of \SI{20}{\milli\meter} diameter made from
\SI{36}{\micro\meter} coated PET film operated at slightly above \SI{1}{\bar}, for a total of 9600 straws
arranged in $y$--$u$--$v$--$y$ planes with a stereo angle of $\sim\SI{5}{\degree}$, for a target
position resolution of \SI{120}{\micro\meter}.

Prototyping of a full-length \SI{4}{\meter} module is underway. The spacing between stations
trades off lever arm against decay-volume length and is being tuned for the best combination of
vertex and momentum resolution. The straw frame must support a structure of this size while
contributing as little material as possible, both to limit multiple scattering and to reduce
electromagnetic debris. Steel, aluminium and carbon fibre are under comparison. Finally, the
magnetic field map is being optimised for reconstruction efficiency across the full range of FIP
models rather than for any single benchmark, and a design decision is expected shortly.

\subsection{Timing detector}

The timing detector consists of three columns of 110 long scintillator bars with a few centimetres
of overlap between columns, each bar read out by 16 SiPMs. Its function is to suppress
combinatorial background. Working together with the UBT, the detector defines strict time windows
around each candidate. The achieved resolution is $\leq\SI{50}{\pico\second}$, with work ongoing
in the optimisation of the SiPMs and their front-end electronics.

\subsection{Calorimeter system}

The calorimeter provides electromagnetic and hadronic energy measurement and the particle
identification used for background rejection. It is based on the SplitCal
concept~\cite{splitcal}: in addition to energy reconstruction with scintillator bars read out by
SiPMs, dedicated high-precision layers inserted at a well-chosen depth measure the direction of
electromagnetic showers. This directionality is what allows fully neutral final states such as
$X\to\gamma\gamma$ to be reconstructed. 

A test-beam campaign concluded in 2025 demonstrated both the particle identification performance
and the angular resolution of the concept, validating the design; the remaining optimisation is
being carried out in simulation. The collaboration is evaluating the reuse of LHCb HCAL modules together with the SPD/PS for the hadronic section, which offers
significant cost savings at no cost in physics performance. For the readout, the choice lies
between the KLauS and CALOROC~B ASICs, for a total channel count between roughly 15\,000 and
96\,000 depending on the final segmentation.

\section{Status, schedule and conclusion}
\label{sec:status}

The facility is in its final design phase and the target design and muon shield layout are essentially frozen, while
background estimates continue to be refined for different operational scenarios and signal
hypotheses. All detector subsystems are in the prototyping phase and are converging rapidly on
their final designs. The Beam Dump Facility is foreseen to be commissioned with beam by 2033; commissioning of the experiment is
planned during the same year, with first results being produced based on data accumulated during
Run~4, allowing SHiP to set world-leading limits over much of its parameter space over a single
year of running.

\acknowledgments

The author was supported by the Research Foundation -- Flanders (FWO) under grant \#12A4O26N and thanks the organisers for the stimulating conference.



\begin{thebibliography}{99}
\setlength{\itemsep}{0pt}

\bibitem{briefingbook}
J.~de~Blas et al., \emph{Physics briefing book: 2026 update of the European Strategy for
Particle Physics}, \href{https://doi.org/10.23731/CYRM-2025-008}{CERN-2025-008}.

\bibitem{physicscase}
S.~Alekhin et al.,
\emph{A facility to search for hidden particles at the CERN SPS: the SHiP physics case},
\href{https://doi.org/10.1088/0034-4885/79/12/124201}
{\emph{Rept.\ Prog.\ Phys.} \textbf{79} (2016) 124201}
[{\tt arXiv:1504.04855}].

\bibitem{proposal}
SHiP Collaboration,
\emph{BDF/SHiP at the ECN3 high-intensity beam facility},
Tech.\ Rep.\ CERN-SPSC-2023-033, SPSC-P-369, CERN, Geneva (2023).

\bibitem{ship2025}
SHiP Collaboration,
\emph{SHiP experiment at the SPS Beam Dump Facility},
{\tt arXiv:2504.06692}.

\bibitem{mushield}
SHiP Collaboration, A.~Akmete et al.,
\emph{The active muon shield in the SHiP experiment},
\href{https://doi.org/10.1088/1748-0221/12/05/P05011}
{\emph{JINST} \textbf{12} (2017) P05011}
[{\tt arXiv:1703.03612}].

\bibitem{fastic}
J.~Mauricio et al.,
\emph{FastIC+: an analog front-end including on-chip TDCs for fast timing detectors},
in proceedings of \emph{2024 IEEE NSS/MIC/RTSD},
\href{https://doi.org/10.1109/NSS/MIC/RTSD57108.2024.10657159}{IEEE (2024)}.

\bibitem{splitcal}
W.~M.~Bonivento,
\emph{Studies for the electromagnetic calorimeter SplitCal for the SHiP experiment at CERN with
shower direction reconstruction capability},
\href{https://doi.org/10.1088/1748-0221/13/02/C02041}
{\emph{JINST} \textbf{13} (2018) C02041}.

\end{thebibliography}
\end{document}